\documentclass[twocolumn,aps,prl,groupedaddress]{revtex4-1}
\usepackage{graphicx}

\newcommand{\xone} {{\mbox{Co${}_3$O${}_4$}}}
\newcommand{\xnul} {{\mbox{CoAl${}_2$O${}_4$}}}
\newcommand{\CoAlO} {{\mbox{Co(Al${}_{1-x}$Co${}_{x}$)${}_2$O${}_4$}}}
\newcommand{\half}{{\ensuremath{\frac{1}{2}}}}
\newcommand{\trhalf}{{\ensuremath{\frac{3}{2}}}}
\newcommand{\four}{{\ensuremath{\frac{1}{4}}}}

\newcommand{\eight}{{\ensuremath{\frac{1}{8}}}}
\newcommand{\qx}{{\ensuremath{\frac{{\bf{q}}_x}{4}}}}
\newcommand{\qy}{{\ensuremath{\frac{{\bf{q}}_y}{4}}}}
\newcommand{\qz}{{\ensuremath{\frac{{\bf{q}}_z}{4}}}}

\begin{document}

\title{Spin liquid in a single crystal of the frustrated diamond lattice antiferromagnet CoAl${}_2$O${}_4$}

\author{O. Zaharko}
\affiliation{Laboratory for Neutron Scattering, Paul Scherrer Institute, CH-5232 Villigen, Switzerland}
\author{N. B. Christensen}
\affiliation{Materials Research Division, Ris\o~National Laboratory for Sustainable Energy, Technical University of Denmark, Denmark}
\author{A. Cervellino}
\affiliation{Swiss Light Source, Paul Scherrer Institute, CH-5232 Villigen, Switzerland}
\author{V. Tsurkan}
\affiliation{Experimental Physics V, Center for Electronics Correlations and Magnetism, University of Augsburg, D-86159 Augsburg, Germany,
Institute of Applied Physics, Academy of Sciences of Moldova, MD-2028 Chisinau, Republic of Moldova}
\author{A. Maljuk}
\affiliation{Leibniz Institute for Solid State and Materials Research Dresden, Helmholtzstrasse 20,
01069 Dresden, Germany}
\author{U. Stuhr, C. Niedermayer}
\affiliation{Laboratory for Neutron Scattering, Paul Scherrer Institute, CH-5232 Villigen, Switzerland }
\author{F. Yokaichiya}
\affiliation{Laborat—rio Nacional de Luz S\'{i}ncrotron, CEP 13083-970 Campinas-SP, Brasil}
\author{D. N. Argyriou}
\affiliation{Helmholtz-Zentrum Berlin f\"{u}r Materialien und Energie, D-14109 Berlin, Germany}
\author{M. Boehm}
\affiliation{Institut Laue-Langevin, 156X, 38042 Grenoble C\'{e}dex, France}
\author{A. Loidl}
\affiliation{Experimental Physics V, Center for Electronics Correlations and Magnetism, University of Augsburg, D-86159 Augsburg, Germany}

\date{\today}

\begin{abstract}
We study the evidence for spin liquid in the frustrated diamond lattice antiferromagnet CoAl${}_2$O${}_4$ by means of single crystal neutron scattering in zero and applied magnetic field.
The magnetically ordered phase appearing below T$_N$=8 K remains nonconventional down to 1.5 K. The magnetic Bragg peaks at the $\bf{q}$=0 positions 
are broad and their lineshapes have strong Lorentzian contributions. Additionally, the peaks are connected by weak diffuse streaks oriented along the $<$111$>$ directions.
The observed short-range magnetic correlations are explained within the spiral spin-liquid model. 
The specific shape of the energy landscape of the system with an extremely flat energy minimum around $\bf{q}$=0 
and many low lying excited spiral states with $\bf{q}$= $<$111$>$ results in thermal population of this manifold at finite temperatures.  
The agreement between the experimental results and the spiral spin-liquid model is only qualitative indicating that microstructure effects might be important to achieve quantitative agreement.
Application of a magnetic field significantly perturbs the spiral spin-liquid correlations. The magnetic peaks remain broad but acquire more Gaussian lineshapes and increase in intensity. The 1.5 K static magnetic moment increases from 1.58 $\mu_B$/Co at zero field to 2.08 $\mu_B$/Co at 10 T. The magnetic excitations appear rather conventional at zero field. 
Analysis using classical spin wave theory yields values of the nearest and next-nearest neighbor exchange parameters J$_1$=0.92(1) meV and J$_2$=0.101(2) meV and an additional anisotropy term D=-0.0089(2) meV for CoAl${}_2$O${}_4$. 
In the presence of a magnetic field, the spin excitations broaden considerably and become nearly featureless at the zone center.
\end{abstract}


\maketitle

\section{Introduction{\label{1}}}
\begin{figure}[tbh]
\includegraphics[width=80mm,keepaspectratio=true]{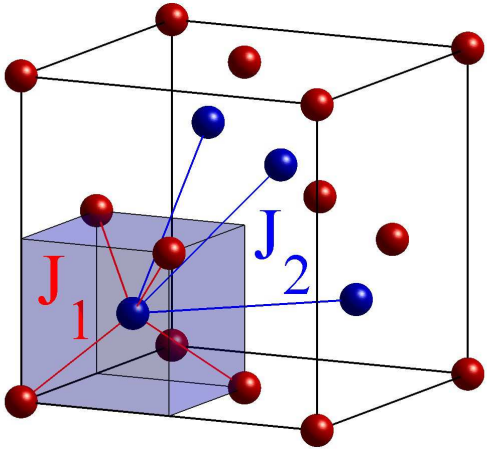}
\caption {The cubic diamond lattice, which can be viewed as two $\it{fcc}$ sublattices shifted by ($\four$ $\four$ $\four$) along $<$111$>$, with the two exchange interactions $J_1$ and $J_2$ indicated.}
\label{fig0}
\end{figure}

Spin liquids, exotic states strongly fluctuating within their degenerate ground states\cite{Balents10}, 
usually form due to the frustrated geometry of the underlying crystalline lattice or due to competing exchange interactions which cannot be satisfied simultaneously. 
The most well-known spin liquid states are realized in materials spanning the pyrochlore, kagom\'{e} or triangular lattices \cite{Gingras09, Ramirez94}. Yet, recently it has been found \cite{Fritsch04, Krimmel05, Tristan05, Bergman07, Bernier08} that even the diamond lattice (Fig.~\ref{fig0}) - despite being bipartite -
can also host a highly degenerate state, the so-called spiral spin liquid. This is a consequence of frustration caused by 
the next-nearest neighbor antiferromagnetic interaction $J_2$, which couples nearest neighbor sites of each FCC sublattice of diamond structure. Nearest neighbor interactions $J_1$ couple sites on different FCC sublattices and can increase the degree of frustration.\\
According to both classical and quantum treatments of this problem\cite{Bergman07, Bernier08} the ground state of the diamond lattice AF is highly degenerate, when the ratio between the couplings,  $J_2/J_1$, exceeds $\eight$. 
The lowest energy degenerate states are spin-spiral states with propagation vectors on a continuous surface in
momentum space. Thermal (entropic) or quantum fluctuations can select a unique ground state from this manifold of degenerate spirals in a process
referred to as Òorder by disorderÓ.
Such exotic physics is realized in the AB$_2$X$_4$ spinels, when the A-site is occupied by magnetic $3d$ ions and the B-site ions are non-magnetic. Large "frustration" ratio (f =$\mid$T$_{CW} \mid$/T$_N \approx$ 10 - 20), specific heat anomalies and liquid-like magnetic  structure factors have been reported for MSc$_2$S$_4$ (M=Mn, Fe) and MAl$_2$O$_4$ (M=Co, Fe, Mn) A-site spinels \cite{Fritsch04, Krimmel05, Krimmel06, Tristan05, Tristan08, Krimmel09}.\\
Recently we studied the evolution of magnetic states in the Co(Al$_{1-x}$Co$_x$)$_2$O$_4$ series by neutron powder diffraction and Monte-Carlo simulations \cite{Zaharko10}. 
The diffraction data indicated that the spin-liquid regime~\cite{note1} emerges within the entire composition range 0$\le x \le$1 at finite temperatures, up to the Curie-Weiss temperature  $\mid$T$_{CW} \mid$=110 K. 
When the composition was changed from $x$=1 to $x$=0, the frustration became stronger, the N\'{e}el temperature T$_N$ decreased and the spin-liquid regime widened.
Comparison of the experimental and the calculated patterns for the Monte-Carlo ground states for various $J_2/J_1$ showed that all compositions in this series belong within the weakly frustrated limit with $J_2/J_1< \eight$ and have conventional antiferromagnetically ordered ground states. 
Our powder diffraction results revealed that the Fourier transform of the radially averaged spin-spin correlations is predominantly Voigtian. However, no further details about the spatial distribution of magnetic correlations and their static or dynamic origin could be extracted from the powder data. 
These limitations can now be overcome due to a recent breakthrough in growing large single crystals of ${\xnul}$~\cite{Maljuk09}.\\
Here we report on single crystal neutron scattering experiments aimed to uncover essential details of the emerging spin liquid. We probe the
spatial distribution of spin correlations, differentiate the static and dynamic contributions, derive the exchange-coupling constants from 
the spin wave dispersion and perturb the spin liquid in ${\xnul}$ by a magnetic field.
We conclude that the observed features can be qualitatively accounted for by the spiral spin-liquid model \cite{Bergman07}, but also that inclusion of microstructure effects might be needed for a quantitative understanding.
 
\section{Experimental}{\label{2}}
The neutron scattering experiments were performed on the cold triple-axis spectrometers TASP and RITA-II at the SINQ spallation source, Villigen, Switzerland and  IN14 at the Institut Laue-Langevin in Grenoble, France. We used a 25 mm long, 7 mm diameter single-crystal of ${\xnul}$ grown by the floating-zone method \cite{Maljuk09} and oriented with the [100] and [011] directions in the horizontal scattering plane. Two setups of TASP were employed: For high energy resolution studies aimed at separating the elastic scattering from inelastic scattering in ${\xnul}$, we used neutrons of fixed final wavenumber $\bf{k}_f$=1.4 \AA$^{-1}$ yielding an elastic energy resolution of 0.1 meV. To study the magnetic excitations, a higher flux setup with $\bf{k}_f$=1.97 \AA$^{-1}$ and a relaxed energy resolution of 0.3 meV was used. RITA-II was operated in the monochromatic imaging mode with seven PG(002) analyzer crystals in front of a position sensitive detector. With 
$\bf{k}_f$=1.55 \AA$^{-1}$ this setup gave an energy resolution of 0.188 meV.  On IN14 we used the FlatCone multianalyzer setup
in which 31 Si(111) analyzers are set to reflect $\bf{k}_f$=1.5 \AA$^{-1}$ neutrons onto detectors located above the horizontal
scattering plane of the spectrometer. This setup allowed to efficiently map out the diffuse elastic scattering of ${\xnul}$ as a function of temperature. 
To study the composition dependence of $J_1$ and $J_2$ in the Co(Al$_{1-x}$Co$_x$)$_2$O$_4$ series, we also performed a short 
experiment on IN14 where the spin wave excitations of an assembly 
of three tiny coaligned crystals (total mass 0.12 g) of ${\xone}$ were studied. In this experiment, the scattering plane was defined by the the [100] 
and [010] axes, and we used a conventional focusing analyser setup with $\bf{k}_f$=1.5 \AA$^{-1}$. For the experiments on TASP and IN14, a standard ILL "orange" cryostat was used to control the sample temperature, whereas on RITA-II we employed a 15 T Oxford Instruments cryomagnet.

\section{Results and discussion}{\label{3}}
\subsection{Zero magnetic field}
\subsubsection{Static spin correlations} 
\begin{figure}[tbh]
\includegraphics[width=90mm,keepaspectratio=true]{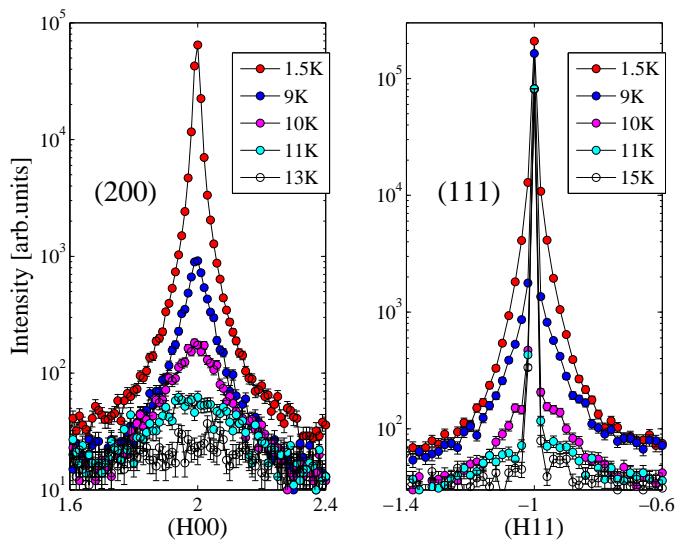}
\caption {Magnetic (200) and mixed nuclear-magnetic (111) peaks at several representative temperatures. The (111) peak with only nuclear contribution at 15 K is a reasonable estimate of the instrumental momentum resolution.}
\label{fig1}
\end{figure}

{\label{Stat}}
The onset of antiferromagnetic order at T$_N$=8 K in ${\xnul}$ is affirmed by the appearance of the purely magnetic reflection (200) (Fig.~\ref{fig1}) rising from a broad diffuse magnetic background and by a substantial increase of the intensity of the mixed nuclear-magnetic (111) reflection. The magnetic contributions to (200) and (111) are clearly much broader than the sharp, resolution limited nuclear Bragg peaks visible $\it{e.g.}$ in the 15 K data at (111). Moreover, within the limitations set by the experimental energy-resolution of 0.1 meV, the magnetic contributions correspond to static spin correlations. It is therefore apparent that the magnetic order emerging below T$_N$ is not a conventional long-range AF order.\\
To quantify the magnetic moment involved in the static correlations, we employed a simple model of an equidomain long-range ordered collinear AF structure with moments pointing along all possible directions of the $<$111$>$ set. The resulting moment at 1.5 K is m=1.58 $\mu_{B}/Co$, which is only half of the saturated moment value of 3 $\mu_B$ expected for Co$^{2+}$ ions. Difference between the properties of the single crystal (T$_N$=8 K, m=1.58 $\mu_{B}$/Co, 8\% site inversion) studied in this work with the polycrystalline sample of Ref.  \cite{Zaharko10} (T$_N$=5 K, m=0.25(7) $\mu_{B}$/Co, 17\% site inversion), probably signifies the importance of $\it{microstructural}$ effects, such as grain boundaries and/or site inversion, in reducing the magnetic ordering temperature and magnetic moment of polycrystalline samples.\\
\begin{figure}[tbh]
\includegraphics[width=90mm,keepaspectratio=true]{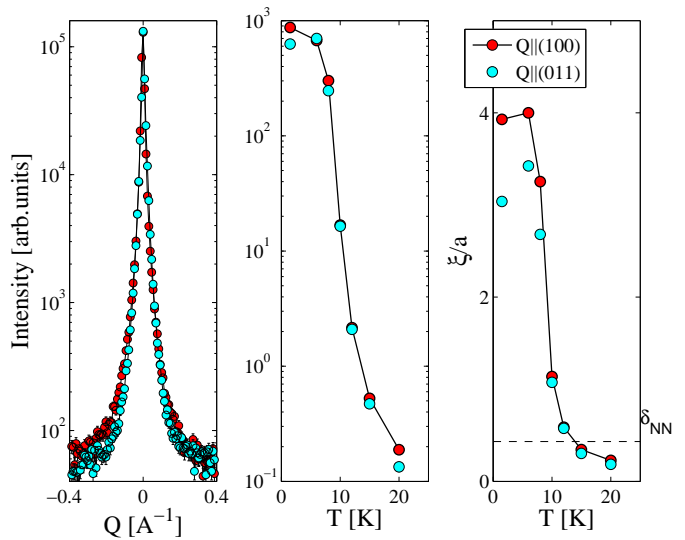}
\caption {Reciprocal space lineshapes of the (200) peak in the [100] and [011]  directions at 1.5 K (left). Temperature evolution of the integrated intensity (middle) and the correlation length (right) in units of the lattice spacing $a$ (a=8.09288 \AA) resulting from fits of the (200) peak to the Pearson VII function with $N=\trhalf$.}
\label{fig2}
\end{figure}
To obtain the length of magnetic correlations from the reciprocal space lineshapes of the broad magnetic contributions
we use the Pearson VII function 
\begin{equation}
P=\frac{1} {(1+\xi^2({\bf{Q}}-{\bf{G}}_M)^2)^N}
\label{eq1}
\end{equation}
with $\xi$ being a measure of the correlation length, ${\bf{Q}}$ - the scattering vector, ${\bf{G}}_M$ - the magnetic lattice vector and $N$ - a real number. This function allows a continuous variation from the pure Lorentzian ($N$ =1) to the pure Gaussian ($N = \infty$) function \cite{Hall77} and it is a good approximation of a Voigt function \cite{Wang05}.\\
The best fits of the lineshapes of the (200) magnetic reflection in zero magnetic field (Fig.~\ref{fig2}) are obtained with the Pearson VII $N= \trhalf$ function. The lineshapes in the transversal [011] and longitudinal [100] directions are similar (Fig.~\ref{fig2} left), suggesting that the spin correlations are spatially isotropic.
The correlation length $\xi$ related to the half-width at half-maximum (HWHM) by the formula $\xi=\left({\sqrt{2^N-1}}\right)/{\mathrm{HWHM}}$, extends over four lattice spacings ($\approx$ 30 \AA) at low temperatures and approaches the nearest neighbour distance $\delta_{NN}$=3.5 \AA~of the A-sublattice near T$_N$ (Fig.~\ref{fig2} right).\\
\begin{figure}[tbh]
\includegraphics[width=80mm,keepaspectratio=true]{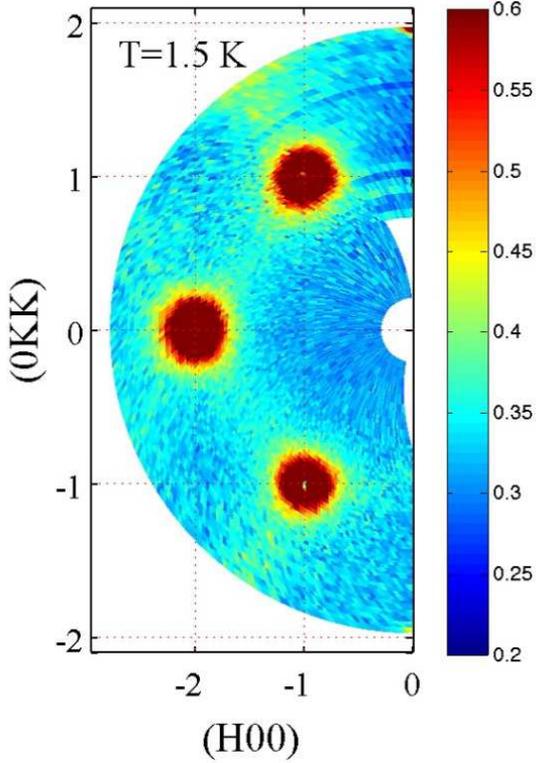}
\caption {Elastic neutron scattering map obtained by division of the 1.5 K data by the 150 K data, as described in the text.}
\label{fig3}
\end{figure}
The broad magnetic peaks are clearly dominating the diffraction pattern but are not the only visible feature of the nonconventional magnetic state in $\xnul$. Careful inspection of the region between reciprocal lattice points reveals that peaks are connected by weak diffuse scattering ('streaks') which is weaker by several orders of magnitude. This scattering is especially well documented in the maps (Fig.~\ref{fig3}) collected  with the FlatCone multianalyzer on IN14. To eliminate the variation in overall scattering due to the strong neutron absorption by the sample, the maps have been processed by dividing the data collected at a certain temperature by the data obtained in the paramagnetic state at 150 K, $\it{i.e.}$ well above T$_{CW}$ where spin liquid correlations set in. Cuts through the streaks perpendicular to the directions joining neighboring broad magnetic peaks were fitted by the Pearson VII function with $N={\trhalf}$ (Fig.~\ref{fig4}), which was also used for  the dominant magnetic contribution. The temperature dependence of the intensity of the magnetic peaks and streaks are very similar, allowing us to attribute both features to the same magnetic ordering phenomenon.\\
\begin{figure}[tbh]
\includegraphics[width=90mm,keepaspectratio=true]{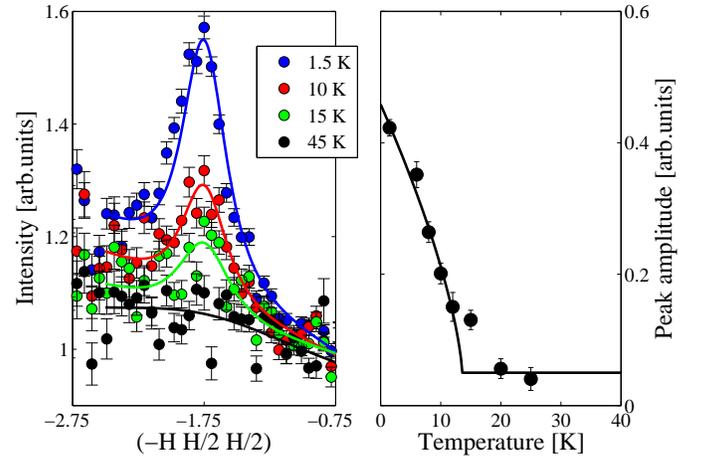}
\caption {Left: Cuts though the diffuse streak at several representative temperatures. Right: Temperature dependence of the corresponding peak amplitude.}
\label{fig4}
\end{figure}
Next, we will demonstrate that both diffuse features, broad peaks and streaks, are consistent with the classical theory of a spiral spin-liquid developed in \cite{Bergman07, Lee08}.
The ground state of the classical Heisenberg antiferromagnet on the diamond lattice can be found considering the Hamiltonian with two dominant AF couplings $J_1$, $J_2$ and the anisotropy constant $D$
\begin{equation}
H = J_{1} \mathop{\sum}_{<ij>} {\bf{S}}_{i} \cdot {\bf{S}}_{j}+ J_{2} \mathop{\sum}_{<<ij>>} {\bf{S}}_{i} \cdot {\bf{S}}_{j} +D\mathop{\sum}_{i}( {\bf{S}}_{i} \cdot{\bf{u}})^2,
\label{eq4}
\end{equation}
${\bf{u}}$ is the anisotropy direction, that was taken to be [111] \cite{note2}. The inclusion of an anisotropy term is motivated by the observation of a gap in the spin wave spectrum to be discussed later. 
The ground state is a spiral defined by the equation
\begin{equation}
{\bf{M}} = {\bf{A}} \cos({\bf{q\cdot r}}+\phi)+{\bf{B}} \sin({\bf{q\cdot r}}+\phi)
\label{eq2}
\end{equation}
where $\bf{M}$ is the magnetic moment at position $\bf{r}$, $\bf{q}$ is the wave vector ($\bf{q}=\bf{Q}-\bf{H}$, with $\bf{H}$ a Bragg peak such that $\bf{q}$ is located in the first Brillouin zone), $A=B=$~1, $\bf{A}\cdot\bf{B}=$~0.
The energy of the ground state is given by
\begin{equation}
E({\bf{q}}) = E_{12}({\bf{q}}) + E_{Anis},
\label{eq5}
\end{equation}
where the contribution $E_{12}$ due to the exchange interactions $J_1$ and $J_2$ is 
\begin{equation}
E_{12} = 16 J_2 \left(\Lambda-\frac{|J_1|}{8J_2}\right)^2 - 4J_2 - \frac{J^2_1}{2J_2}
\label{eq6}
\end{equation}
with the reciprocal space variable
\begin{equation}
\Lambda({\bf{q}}) = \left[\cos^2 \qx \cos^2 \qy \cos^2 \qz + \sin^2 \qx \sin^2 \qy \sin^2 \qz \right]^{\half},
\label{eq7}
\end{equation}
while the contribution $E_{Anis}$ of the anisotropy term $D$ is
\begin{equation}
E_{Anis} = \frac{D}{2} (1-\cos^2{\alpha})
\label{eq8}
\end{equation}
where $\alpha$ is the angle between the anisotropy axis ${\bf{u}}$ and $\bf{A}\times \bf{B}$.
This result is similar to the one derived in Ref.~\cite{Lee08}, with the difference that the third-neighbor coupling $J_3$ in Eq. 8 of ~\cite{Lee08} is replaced by the anisotropy term $D$.\\
For each orientation of a spiral at each ${\bf{Q}}$-point we need to calculate the energy $E({\bf{q}})$ and the probability of this state $P\propto {\mathrm{e}}^ \frac{-E({\bf{q}})}{kT}$. The intensity of 
each populated spiral has to be averaged over the 'full phase' $\omega={\bf{q\cdot r}}+\phi$ 
\begin{equation}
<Z>=\frac{1}{2\pi} \int\limits_{0}^{2\pi}d\omega({\bf{M}}_{\perp})^2 =\half(1+\cos^2\beta)
\label{eq3}
\end{equation}
where ${\bf{M}}_{\perp}$ is the component perpendicular to ${\bf{Q}}$ of the magnetic moment $\bf{M}$ and $\beta$ is the angle between $\bf{q}$ and the vector $\bf{A}\times \bf{B}$.
Summation of the $<Z>$ contributions of all spirals with the probability $P$ gives a good approximation of the total intensity of such a system at a given temperature.\\
Fig.~\ref{fig5} shows diffraction patterns calculated for the ($H00$)-($0KK$) plane following this procedure for two different temperatures.
In the left frame of Figure~\ref{fig5}, the pattern is calculated at a temperature T/$J_1$=0.01 using the experimentally determined exchange parameters for $\xnul$  $J_1$=0.92(1) meV, $J_2$=0.101(2) meV (see the following section). The ratio $J_2/J_1=0.109(2)$ of the exchange parameters and the equivalent temperature of the physical system, T=0.1 K -- well below the base temperature of the experiment -- implies that our calculation probes the theoretical ground state, which is the conventional long-range ordered two-sublattice antiferromagnet \cite{Bergman07}. Indeed, it is clear from the left panel of Figure~\ref{fig5}  that one expects sharp Bragg peaks for T/$J_1$=0.01. 
In Figure~\ref{fig5} right, the diffraction pattern is calculated for T/$J_1$=0.187. With the experimentally determined exchange parameters, this corresponds to a temperature T=2 K of $\xnul$, just above the experimental base temperature. In this ($H00$)-($0KK$) map the Bragg peaks are much broader than
at T/$J_1$=0.01 and diffuse 'streaks' between the magnetic peaks develop.\\
We now turn to the origin of broad Bragg peaks and diffuse streaks in the theoretical calculation performed for T/$J_1$=0.187. The Bragg peaks are broad due to an extremely flat energy minimum around $\bf{q}$=0, $\it{i.e.}$ there exists a large number of low lying excited states, which are populated at this finite temperature. Similarly, the diffuse streaks arise from thermal population of low-energy co-planar spiral states with propagation vectors parallel to $<$111$>$. Indeed, for $J_2$/$J_1<\eight$, these states have lower energies than excited states with propagation vectors along other symmetry directions. (For $\eight < J_2/J_1 \le \four$ entropy even selects spirals states with propagation vectors along $<$111$>$ as the ground state, due to the higher density of nearby low-energy states~\cite{Bergman07}.)\\
While the above comparison reveals important similarities
between the experimentally observed and theoretically computed diffraction patterns, it should be noted that the agreement is qualitative: The calculated lineshapes are Gaussian, while the experimental lineshapes display a significant Lorentzian component. In addition, the temperature dependences of the calculated intensity for the 'peak' and the 'streak' differ below T$_N$ (Fig.~\ref{fig6}), while in the experiment they reveal a similar evolution (Figs.~\ref{fig2} and \ref{fig4}). 
The most natural assumption concerning the origin of these discrepancies is that microstructure effects, such as domain boundaries or lattice defects caused by inversion, interrupt the development of the true long-range order reflected by the diverging peak intensity in Fig.~\ref{fig6}. Instead, the Bragg peaks remain broad and the streaks retain finite intensity, $\it{i.e.}$ the spin-spin correlations remain short-ranged even at the lowest temperatures of our experiment.
\begin{figure}[tbh]
\includegraphics[width=36mm,keepaspectratio=true]{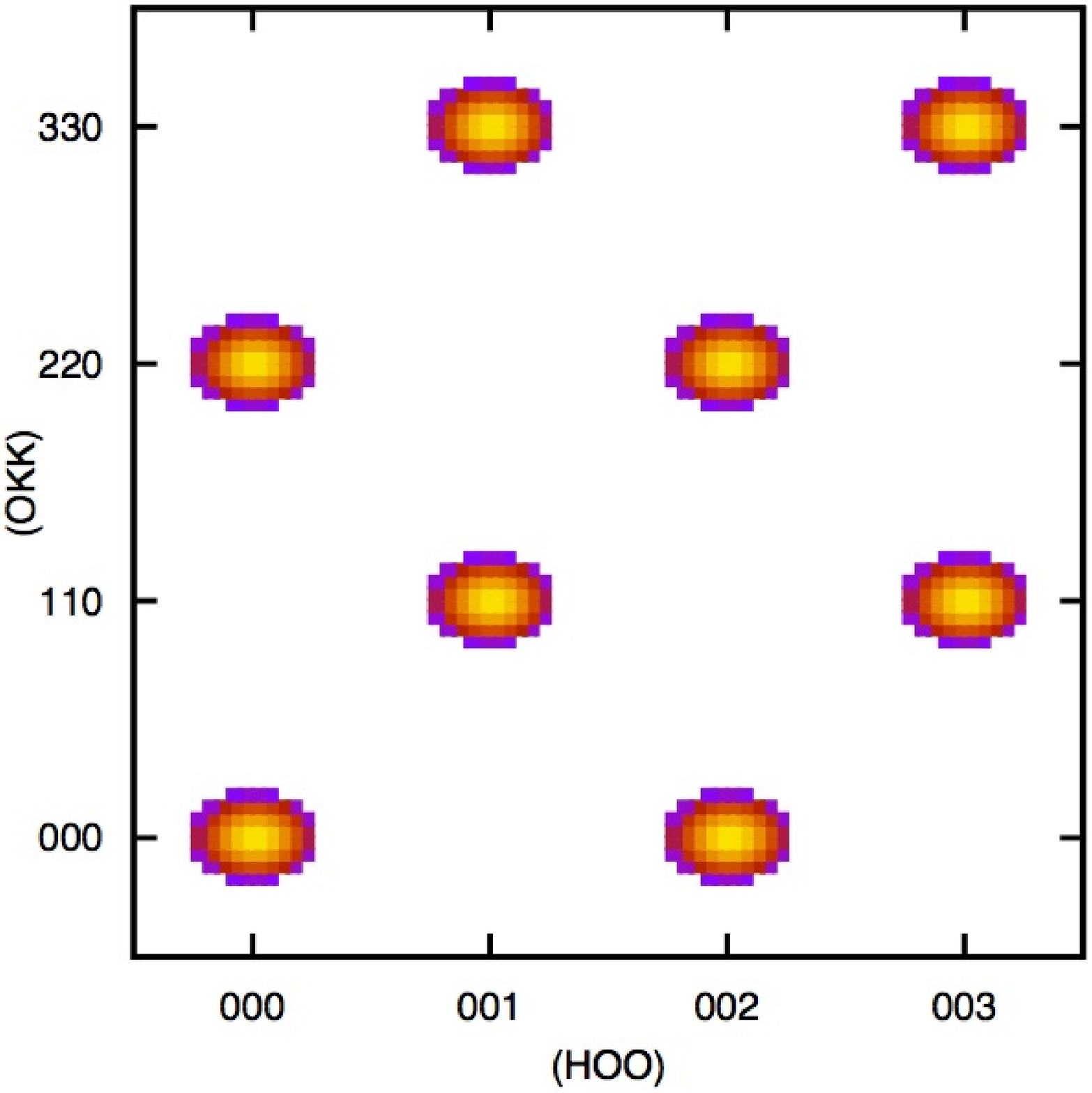}
\includegraphics[width=40mm,keepaspectratio=true]{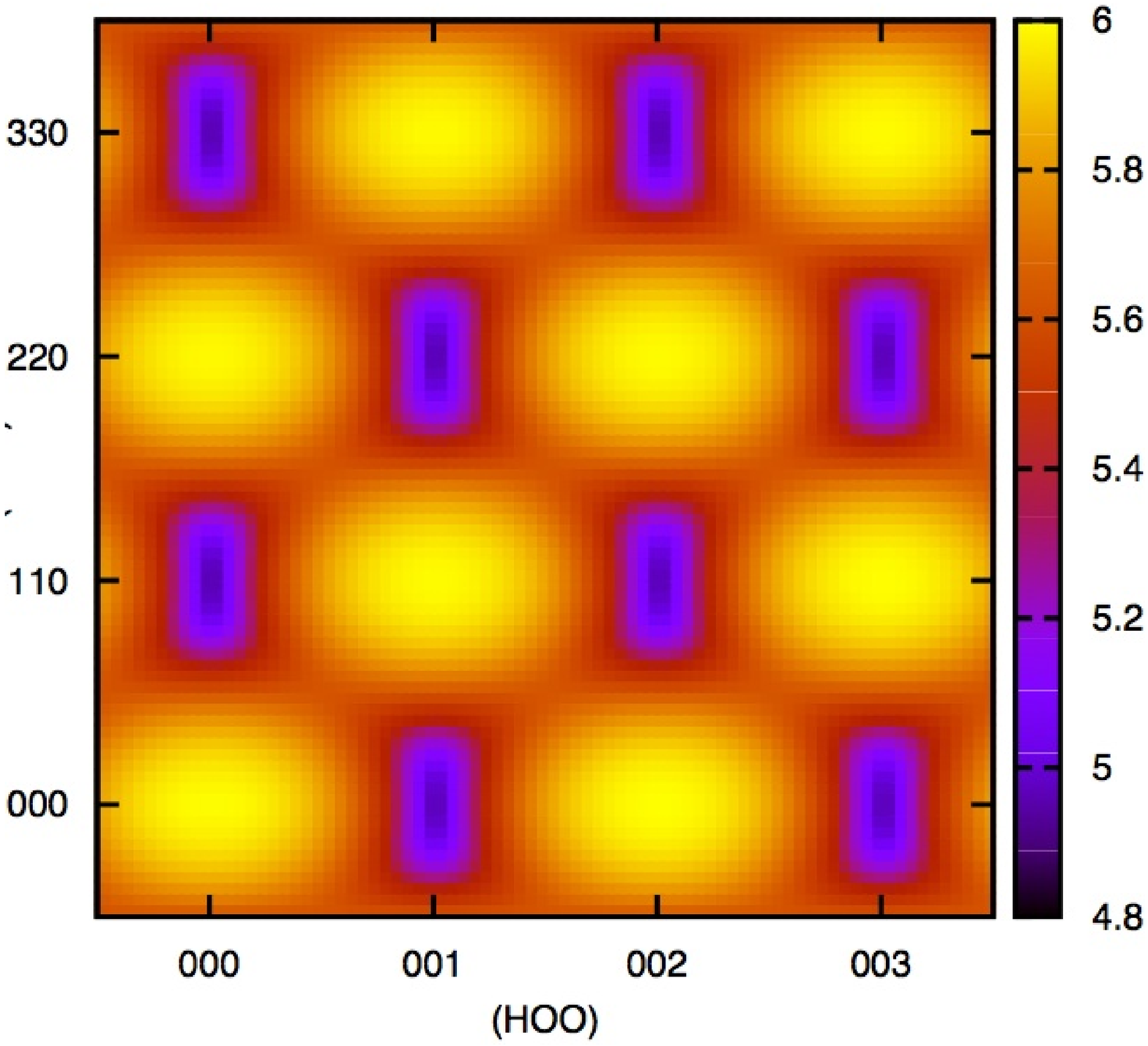}
\caption {Elastic neutron scattering maps calculated with $J_2/J_1$=0.1, $D/J_1$=0.01 at temperatures 0.1 K (left) and 2 K (right). Note the logarithmic scale for the intensity.}
\label{fig5}
\end{figure}
\begin{figure}[tbh]
\includegraphics[width=80mm,keepaspectratio=true]{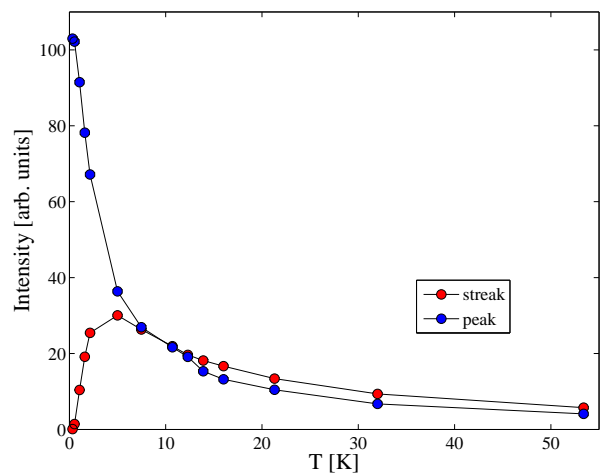}
\caption {Calculated temperature dependence of the intensity of the "peak" and the "streak" based on the spiral spin-liquid model described in the text.}
\label{fig6}
\end{figure}

\subsubsection{Dynamic spin correlations}
\begin{figure}[tbh]
\includegraphics[width=90mm,keepaspectratio=true]{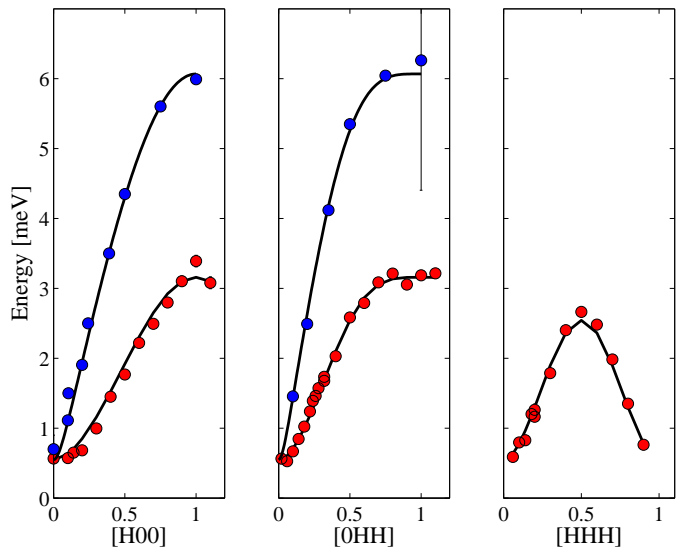}
\caption {Dispersion relations  for the three main directions ([H00], [HH0] and [HHH] are reduced wavevectors) measured at 1.5 K and fitted using a linear spin wave model based on the Hamiltonian of Eq.~\ref{eq4}. Red symbols are for $\xnul$ and blue for $\xone$.}
\label{fig7}
\end{figure}

\begin{figure}[tbh]
\includegraphics[width=90mm,keepaspectratio=true]{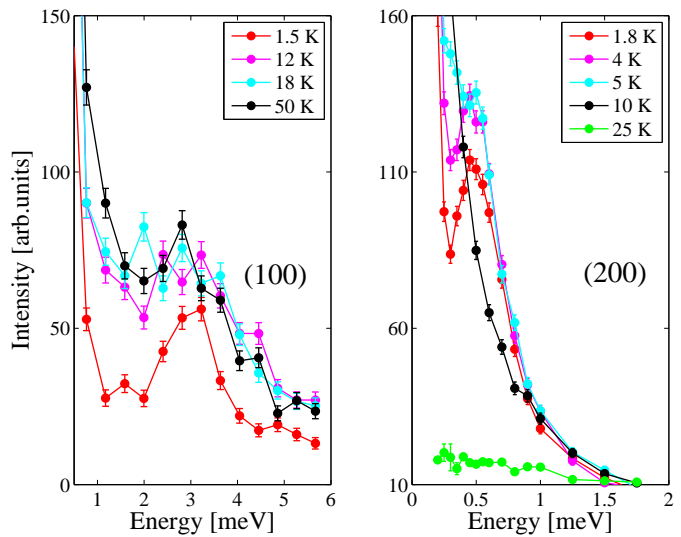}
\caption {Temperature evolution of the spin excitations near the zone boundary (100) (TASP) and of the gap near the zone center (200) (IN14) for $\xnul$. Note, the data are from different instruments and therefore the intensities cannot be compared.}
\label{fig8}
\end{figure}

{\label{Dyn}}
To investigate the dynamic correlations in the low-temperature phase of $\xnul$ and to determine the effective exchange coupling constants, we measured the dispersion relations of spin excitations propagating along the three principal high-symmetry directions [100], [110] and [111] (Fig.~\ref{fig7}).
The gross features of the excitation spectrum look rather conventional at the base temperature of 1.5 K: The width of the excitations are resolution limited in energy, and there is a small gap of the order of 0.5 meV at the magnetic zone center. At the zone boundaries the excitation energy reaches 2.5 meV along the [111] direction and 3.2 meV for the [100] and [110] directions.\\ 
As the temperature is increased towards T$_N$, the zone center gap closes, as expected for a spin anisotropy gap (Fig.~\ref{fig8} right). At temperatures of the order of 3T$_N$ only weak quasielastic scattering remains. By contrast, at the zone boundary wavevectors a strong quasielastic signal persist to at least 90 K (Fig.~\ref{fig8} left).\\
Applying classical spin wave theory to the Hamiltonian defined in Eq.~\ref{eq4} under the assumption of a conventional AF collinear ground state
we determined the parameters $J_1$=0.92(1) meV, $J_2$=0.101(2) meV and $D$=-0.0089(2) meV for $\xnul$ at T=1.5 K. 
This results in a ratio $J_2$/$J_1$= 0.109(2)  of exchange parameters and a Curie-Weiss temperature $\mid$T$_{CW}\mid=(4J_1+12J_2)S(S+1)/3k_B$=105.7 K.
For $\xone$ the same treatment of spin excitations propagating along the [100] and [011] directions (Fig.~\ref{fig7}) leads to
$J_1$=1.09(15) meV, $J_2$=0.02(4) meV and $D$=-0.0076(14) meV. Thus, for $\xone$, $J_2$/$J_1$=0.019(35) and $\mid$T$_{CW} \mid$=99.55 K.\\
Note that the spin wave analysis produces effective exchange parameters that should be treated with caution. 
In fact our diffraction data obtained at the experimental base temperature of 1.5 K clearly show a ground state which is not conventional. 
Nevertheless, using these parameters, we are able to reproduce nearly composition-independent Curie-Weiss temperatures not far from the previously reported value of 110 K \cite{Tristan08}, as well as a significantly larger degree of frustration as quantified by $J_2/J_1$ in $\xnul$ than in $\xone$ \cite{Tristan08,Zaharko10}.
\subsection{Magnetic field}
\subsubsection{Static spin correlations}
\begin{figure}[tbh]
\includegraphics[width=90mm,keepaspectratio=true]{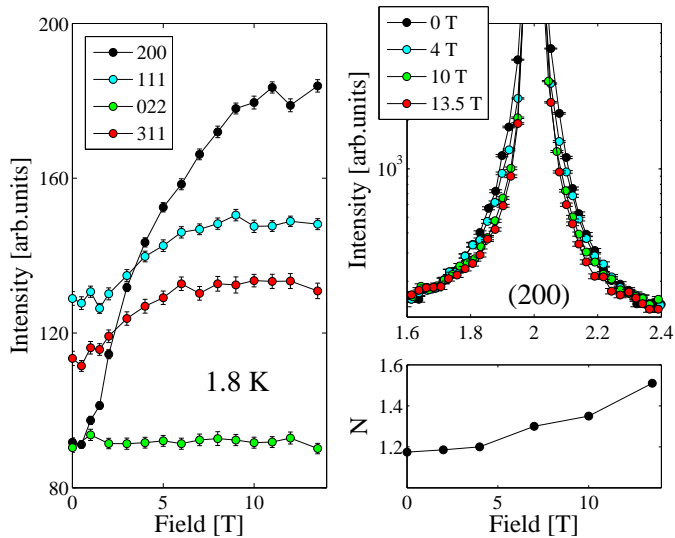}
\caption {Left: Change of the integrated intensities of four selected peaks with applied magnetic field at 1.8 K for CoAl$_2$O$_4$. Right: Evolution of the peak lineshape and the order N of the Pearson VII function used to fit the (200) magnetic reflection.}
\label{fig9}
\end{figure}

{\label{StatH}}
Application of a magnetic field can assist in selecting a certain ordered state in magnetically frustrated systems \cite{Helton07, Zhou07}. We were therefore interested in the effect of a magnetic field on the spiral spin liquid correlations that we observe in $\xnul$.
For the long-range ordered AF ground state expected when $J_2/J_1<\eight$, the theory of Lee and Balents \cite{Lee08} predicts
a saturation field $H_C=8J_1$. Using the effective exchange couplings determined in this work, the saturation field of $\xnul$ is expected to be $H_C=8J_1\approx$ 56 Tesla. 
Our observations, documented below, indicate that fields up to 15 T $\simeq H_C/4$ perturb the spin liquid correlations below $T_N$ substantially. 
The system does not reach the conventional truly long-range ordered AF N\'{e}el state, but the decay of the spin correlations changes significantly.\\
The evolution of the intensities of four selected reflections with magnetic field applied along the [01-1] direction is presented in the left frame of Fig.~\ref{fig9}. The intensity of the nuclear (022) reflection does not change with field indicating that no ferromagnetic component, $\it{e.g.}$ due to canting, is induced at these fields. In addition, the ordering temperature does not change.
The most pronounced observation is that the intensities of the purely magnetic (200) and mixed nuclear-magnetic (111), (311) reflections increase with applied field. Employing a simple model of a single domain AF collinear structure with the magnetic moments orthogonal to the applied field we obtain the moment value of 2.08 $\mu_B$/Co at 1.7 K and H= 10 T. This is 15\% larger than the zero-field value, but still significantly reduced compared to the free-ion value.\\
Interestingly, the lineshapes of the magnetic peaks change notably with applied field (Fig.~\ref{fig9} right up) and the exponent $N$ of the Pearson VII function grows. This, however, should not be misinterpreted as the development of long-range order since the fitted lineshape remains much broader than dictated by the instrumental momentum resolution in the entire field range probed. The increase of $N$ indicates that the decay of static spin correlations changes and approaches a more Gaussian distribution, while the spin correlation length remains short, and the ordered state induced by magnetic field stays unconventional. Within the spiral spin-liquid picture we presume that the applied magnetic field changes the shape of the energy landscape around the theoretical ordering vector $\bf{q}$=0, making it more anharmonic and thus modifying the decay of the spin correlations.
\begin{figure}[tbh]
\includegraphics[width=90mm,keepaspectratio=true]{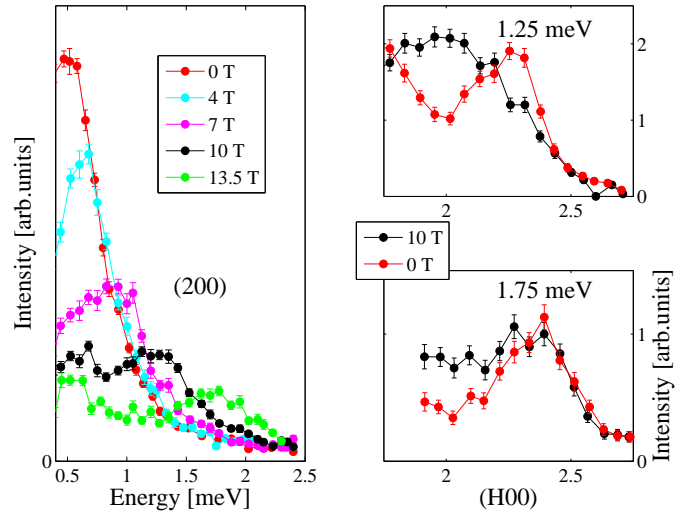}
\caption {Left: Field dependence of the spectrum at the zone center (200) for $\xnul$. Right panels: Constant energy scans at 1.25 meV and 1.75 meV, showing how the magnetic field adds spectral weight at the zone center.
All data shown were taken at 1.8 K.}
\label{fig10}
\end{figure}
\subsubsection{Dynamic spin correlations}{\label{DynH}}

Just as the static spin correlations are affected by an applied magnetic field, the spin excitation spectrum also changes significantly. The changes are especially strong near the zone center (200) (Fig.~\ref{fig10}). 
By contrast, the energy of the spin excitations at the zone boundary wavevector (300) remains unchanged (not shown). 
As the magnetic field is increased, the excitation near (200) centered at 0.5 meV gradually looses intensity, and broadens while moving to higher energies.
The broadening, which is already very significant at 7 T continues to the highest
 investigated field (13.5 T). At this field, a broad band of magnetic intensity is present up to at least 2 meV. The right panel of Figure~\ref{fig10} highlights another aspect of the data: The magnetic field adds spectral weight at the zone center (200), causing a loss of definition of the spin wave modes, which were clearly separated in the 0 T data set.\\
It is tempting to interpret these observations as a splitting of modes which were nearly degenerate in zero-field state. Our instrumental energy resolution was, however, insufficient to clearly resolve the individual modes to 
directly verify this interpretation. Nevertheless, the width of the broad band of magnetic excitations is significantly lower then the Zeeman splitting $g \mu_B H S_z$=4.7 meV expected for noninteracting Co$^{2+}$ moments at 13.5 T, indicating that the excitations remain collective. In a conventional antiferromagnet the magnetic excitations would split when a magnetic field is applied, but would also stay sharp. Therefore the observed continuum-like broad band of excitations supports the idea of a manifold of low-lying spiral-spin states in $\xnul$ with their excitations split by magnetic field.
\section{Concluding remarks}

Our single crystal neutron scattering study brings new insights into the spin-liquid state exhibited by frustrated diamond lattice antiferromagnetic spinels.
We confirm that the system ${\CoAlO}$ is weakly frustrated \cite{Zaharko10}, with the ratio of the exchange coupling constants $J_1$ and  $J_2$ equal to $J_2/J_1$=0.109(2) for $\xnul$ and $J_2/J_1$=0.019(35) for $\xone$. 
Furthermore, our results demonstrate that the Curie-Weiss temperature remains constant in the whole range of compositions 0$\le x \le$1 \cite{Tristan05, Tristan08}  due to compensation between
 the rise of $J_1$ and the decrease of $J_2$ with increasing $x$.\\
Spin-liquid features are pronounced at the finite temperatures of our neutron scattering experiments. In the elastic scattering channel, the spin liquid correlations are seen as atypically broad magnetic Bragg peaks with a Pearson VII line shape (intermediate between Gaussian and Lorentzian) and as weak streaks along the $<$111$>$ directions connecting the peaks. 
These observations are well explained by the spiral spin-liquid model. 
Although the theoretical ground state for $J_2/J_1< \eight$ is the collinear two-sublattice antiferromagnet, the energy minimum around $\bf{q}$=0 is very flat and many low lying excited spiral states are thermally populated even at the lowest temperatures achieved in our experiments, leading to broadening of the magnetic Bragg peaks. 
Likewise, the weak streaks indicate that the $\bf{q}$=$<$111$>$  directions are special for the $J_2/J_1$ ratio of $\xnul$. Indeed, these are the directions with lowest-energy highest-density excited states favored by entropy \cite{Bergman07}.\\ 
The spiral spin-liquid picture, however, is not sufficient to explain all our findings. This is not surprising, as magnetic ordering in frustrated systems is known to be highly susceptible to perturbations. The ground state degeneracy can be broken by site disorder, dipolar or Dzyaloshinskii-Moriya interactions, spin-lattice coupling, etc. \cite{Gingras09, Lee07, Radaelli05, Dutton11}. From our experience with spinel oxides and comparing the published results \cite{Tristan05, Tristan08, Krimmel09, Zaharko10, Maljuk09, MacDougall11}, it is very likely that $\xnul$ is susceptible to microstructure effects. The idea of frozen domain walls invoked to explain the diamond-lattice antiferromagnet $\xnul$ by MacDougall \cite{MacDougall11}, therefore, deserves special attention. Although formation of sharp domain walls in this material characterized by weak anisotropy is unlikely, domain boundaries generated by microstructure are possible. Pinning can be accomplished by nonmagnetic defects on the A-sites or/and by magnetic defects on the B-sites, which exist due to the site inversion and can persist down to the lowest temperatures of the experiment. A mechanism of breaking the degeneracy of the ground state by dilute impurities was proposed by Savary \cite{Savary11}. In their model spin-spiral states deform locally around defects, with the spiral wave vector being a compromise with respect to different impurities.\\
We find that applied magnetic field modifies both the static and the dynamic spin correlations. The magnitude of the ordered moment increases and the decay of the spin correlations changes towards a more Gaussian distribution, suggesting a more strict selection of the occupied states in an applied field. The spin excitations are split by the magnetic field, resulting in a continuum-like spectrum near the ordering wavevector -- presumably due to splitting of states which are nearly degenerate in the zero field state. This effect indirectly confirms the existence of multitude of degenerate states. On the other hand, it significantly complicates any attempt at a quantitative description using conventional spin wave analysis.\\
Further theoretical studies are required to understand the spiral spin liquid in $\xnul$. It is  still not clear what is the role of microstructure effects in the experimental observations in zero and applied magnetic field presented in this work and that by MacDougall \cite{MacDougall11}. Also proposals how to distinguish experimentally the spiral spin-liquid model and the frozen domain-wall model are awaited.\\
\begin{acknowledgments}
This work was performed at SINQ, Paul Scherrer Institute, Villigen, Switzerland and at the ILL reactor, Grenoble, France. The work was supported by the Danish Research Council through DanScatt and by the DFG-TRR80 (Augsburg-Munich).
\end{acknowledgments}


\begin{thebibliography}{99}
\bibitem{note1} Here we use the same definition of the spin liquid as in Ref. \cite{Zaharko10}: Spin-liquid is characterized by rapid (at least exponential) decay of spin-spin correlations and the correlation length does not exceed the interatomic distance\cite{Canals98}, which result in a spin liquid-like structure factor.
\bibitem{note2} This is the easy direction according to the Monte Carlo simulations of Refs. \cite{Bergman07, Zaharko10}, no experimental determination has been done so far.
\end{thebibliography}
\end{document}